\documentclass[aps,preprint]{revtex4}
\usepackage{ifpdf}
\ifpdf
\usepackage[pdftex]{graphicx}
\usepackage{epstopdf}
\else
\usepackage{graphicx}
\fi

\usepackage[nolist]{acronym}

\begin{acronym}
\acro{BEEM}{ballistic electron emission microscopy}
\acro{BHEM}{ballistic hole emission microscopy}
\acro{STM}{scanning tunneling microscopy}
\acro{TEM-EDX}{Transmission electron microscopy energy dispersive x-ray
spectroscopy}
\acro{IV}{I-V}
\acro{TEM}{transmission electron microscopy}
\acro{EDX}{energy dispersive x-ray spectroscopy}
\acro{AFM}{atomic force microscopy}
\acro{AES}{auger electron spectroscopy}
\acro{RBS}{Rutherford backscattering spectrometry}
\acro{SB}{Schottky barrier}
\acro{FLT}{Fermi liquid theory}
\acro{UHV}{ultra high vacuum}
\acro{IBS}{interface band structure}
\acro{MBE}{Molecular Beam Epitaxy}
\acro{BK}{Bell and Kaiser}
\acro{PL}{Prietsch Ludeke}
\end{acronym}

\begin{document}

\title{
Time dependent changes in Schottky barrier mapping of the W/Si(001) interface
utilizing Ballistic Electron Emission Microscopy
}

\author{
Chris A. Durcan
}
\affiliation{College of Nanoscale Science and Engineering, State University of
New York, Albany, New York 12203, USA}

\author{
Robert Balsano
}
\affiliation{College of Nanoscale Science and Engineering, State University of
New York, Albany, New York 12203, USA}

\author{
Vincent P.~LaBella
}\email{vlabella@albany.edu}
\affiliation{College of Nanoscale Science and Engineering, State University of
New York, Albany, New York 12203, USA}

\date{\today}

\begin{abstract}
The W/Si(001) Schottky barrier height is mapped to nanoscale dimensions using
\ac{BEEM} over a period of 21 days to observe changes in the interface
electrostatics. Initially the average spectrum is fit to a Schottky barrier
height of 0.71~eV and the map is uniform with 98$\%$ of the spectra able to be
fit. After 21 days, the average spectrum is fit to a Schottky barrier height of
0.62~eV and the spatial map changes dramatically with only 27$\%$ of the spectra
able to be fit. Transmission electron microscopy shows the formation of an
ultra-thin tungsten silicide at the interface which increases in thickness over
the 21 days. This increase is attributed to an increase in electron scattering
and the changes are observed in the \ac{BEEM} measurements. Interestingly,
little to no change is observed in the I-V measurements throughout the 21 day
period.

\end{abstract}



\maketitle
\acresetall
\section{Introduction}

Schottky diodes are an active area of research due to their low capacitance,
recovery times, and importance in silicon based MOSFET's as source-drain
contacts
~\cite{coss:jvstb:31:021202,smith:mj:82:261,haimsona:me:92:134,papadatos:me:92:2012,tsai:jes:128:2207}.
A Schottky barrier is formed at the interface between a metal and semiconductor.
The chemical potential difference between the two materials causes a transfer of
electrons creating an electrostatic energy barrier called the Schottky
barrier~\cite{cheung:apl:49:85,arizumi:jjap:8:749}. The height of the Schottky
barrier can fluctuate laterally across the interface due to chemical and
physical variations at the metal-semiconductor
interface~\cite{tung:apl:58:2821}. The Schottky-Mott model predicts the Schottky
barrier height is the difference between the Fermi level of the metal and the
electron affinity of the semiconductor. Typically this is not observed
experimentally due to effects such as metal induced gap states, defects, and
semiconductor surface states~\cite{tung:prb:45:13509,
palm:prl:71:2224,fowell:jvstb:9:581, fowell:semst:5:348,olbrich:jap:83:358}. To
capture the nanoscale fluctuations in Schottky barrier height, Tung introduced a
chemical bonding model which predicts the varying bonding angles between metal
and semiconductor will produce a Gaussian like spread in barrier
heights~\cite{tung:prl:84:6078}.

An ideal technique to study Schottky barriers at the nanoscale is \ac{BEEM}. In
\ac{BEEM}, an \ac{STM} tip is used to inject hot electrons into the metal layer
of a Schottky diode while the current transmitted into the semiconductor is
recorded. If injected electrons travel through the metal film without
scattering; and if the electrons carry enough perpendicular momentum to overcome
the Schottky barrier the electrons are collected as \ac{BEEM} current. \ac{BEEM}
probes the interfacial electrostatics of a metal-semiconductor junction by
varying the injected carrier energy and tip location with meV energy resolution
($\sim$0.02~eV) and nanometer spatial resolution, respectively
\cite{prietsch:prep:253:163}. This technique has been used to measure the
Schottky barrier height of many different metal/semiconductor
interfaces~\cite{weilmeier:prb:59:R2521, weilmeier:prb:61:7161,
kaiser:prb:44:6546, sirringhaus:prb:53:15944, bobisch:prl:102:136807,
ventrice:prb:53:3952, sitnitsky:jvstb:30:04E110, garramone:apl:96:062105,
im:apl:72:839, tivarus:prl:94:206803, banerjee:ietm:41:2642,
garramone:apl:100:252102, garramone:jvsta:28:643, garramone:jvstb:27:2044,
stollenwerk:jvstb:24:2009, stollenwerk:jvsta:24:1610, qin:jap:111:013701,
forment:semst:16:975}. H.-J. Im et al. took $\sim$800 spectra of the Pd/SiC
diode and showed a gaussian distribution of Schottky barrier heights consistent
with the Tung model~\cite{im:prb:64:075310}. Similarly for the W/n-Si(001)
interface, a Schottky barrier of 0.71~eV was measured with a histogram of
Schottky barriers obtained from 7225 spectra showed a gaussian distribution with
some higher barrier heights attributed to defects or contaminates at the
interface~\cite{durcan:jap:116:023705}.

In this article, the W/Si(001) Schottky junction is mapped to nanoscale
dimensions over 21 days where every 7 days two point \textit{in situ} \ac{IV}
measurements and \ac{BEEM} Schottky barrier maps were acquired. The distribution
of Schottky barrier heights of the as deposited tungsten film follow a Gaussian
like distribution with few areas of no measured barrier height. Over the 21 day
period the percentage of spectra able to be fit decreased from 98$\%$ to 27$\%$.
\ac{TEM-EDX} shows that a tungsten silicide forms upon deposition which grows in
thickness and changes in composition over time and attributed to the changes
observed in the BEEM data. Interestingly, two point \ac{IV} measurements show
little change over the 21 day period.

\section{Experimental}

The diode fabrication was performed under high-vacuum conditions on lightly
doped $n$-Si(001) substrates with resistivity of 100 $\Omega$-cm. The native
silicon oxide layer was removed utilizing a standard chemical hydrofluoric acid
treatment immediately prior to loading into a high vacuum ($10^{-8}$~mbar)
deposition chamber~\cite{garramone:apl:96:062105,garramone:jvsta:28:643}. The
tungsten was deposited onto the substrate using electron beam evaporation; the
substrate was partially covered using a shadow mask creating diodes 2~mm by
2~mm. The tungsten thickness for the sample was 5~nm. A 10~nm gold layer was
deposited on top of the tungsten layer to inhibit tungsten surface oxidation.
The silicon substrate was unintentionally heated above room temperature due to
radiative heating from the deposition source. The deposition thickness was
calibrated using \ac{RBS} and \ac{TEM-EDX} in a JEOL Titan. After deposition,
the sample was mounted onto a custom designed sample holder for \ac{BEEM}
measurements. The plate allows for simultaneous grounding of both the metal film
and the silicon substrate to an \textit{ex situ} pico-ammeter to measure
\ac{BEEM} current. An ohmic contact was established by cold pressing indium onto
the backside of the silicon substrate.

\begin{figure}
\includegraphics{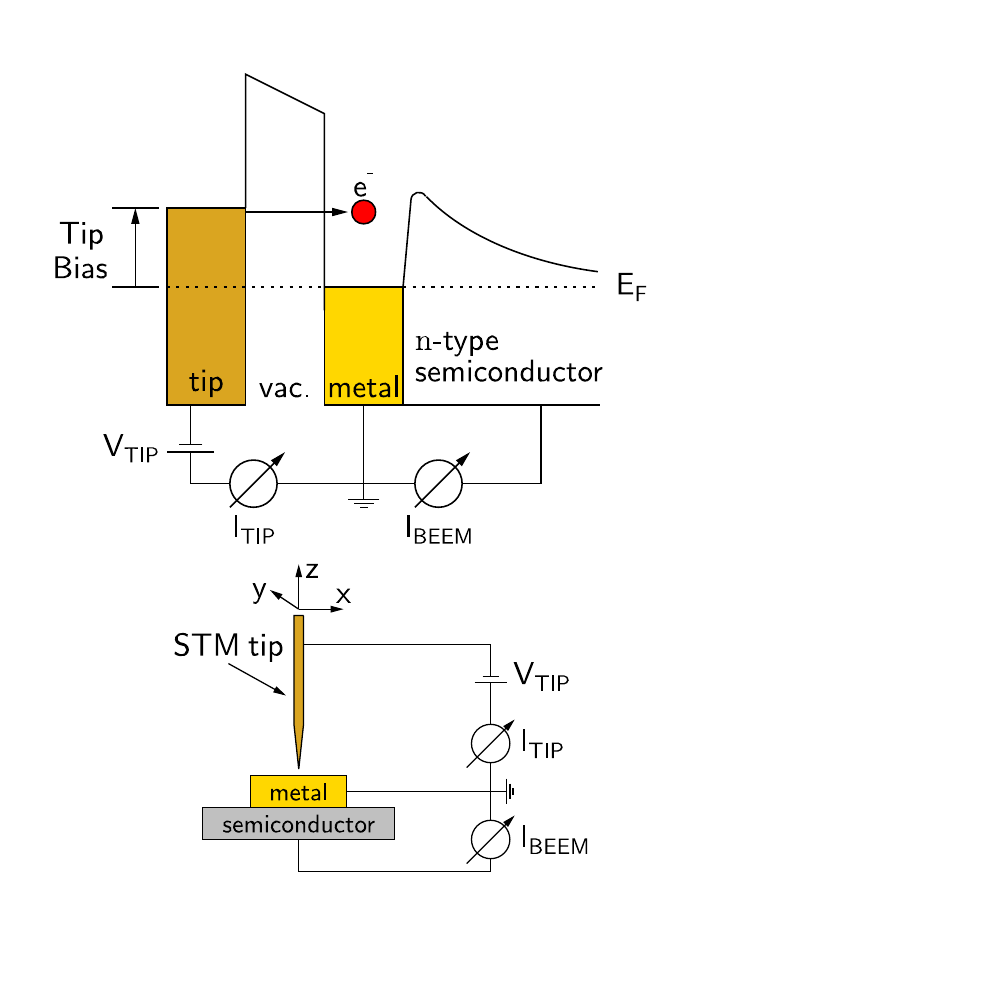}
\centering
\caption{(color online) (top) Energy diagram of the \ac{BEEM} setup with STM
tip, metal and n-type semiconductor. (bottom) Wiring schematic of the \ac{BEEM}
setup.}
\label{schematic}
\end{figure}

A modified low temperature \ac{STM} (Omicron) was utilized for all \ac{BEEM}
measurements operating at a pressure in the $10^{-11}$ mbar
range~\cite{krause:jvstb:23:1684}. The sample was inserted into the UHV chamber
within 3 hours of metal deposition and loaded onto the STM stage and cooled to
80~K for all measurements. Two-point \ac{IV} measurements were taken \textit{in
situ} for each sample at low temperatures using a Keithley 2400 source
measurement unit to record \ac{IV} characteristics. All measurements were taken
in the absence of ambient light. Pt/Ir STM tips, mechanically cut at a steep
angle, were utilized for all BEEM measurements. The experimental schematic and
energy band diagram are shown in Fig.~\ref{schematic}. All \ac{BEEM}
measurements were taken using a constant tunneling current set-point of 20~nA
with forward tip biasing, i.e. the carriers injected into the metal are
electrons, the same as the majority carriers in the silicon substrate.

To obtain a \ac{BEEM} spectrum, the \ac{STM} tip stays at one location while the
voltage between the tip and metal surface is increased from 0.2~V to 2.0~V while
recording the \ac{BEEM} current. The \ac{STM} tip moves to a new location and
the measurement is repeated. For each series of \ac{BEEM} measurements, spectra
were taken every 11.7~nm throughout a 1~$\mu$m~$\times$~1~$\mu$m area. After the
initial set of spectra was taken, the sample was allowed to return to room
temperature and sit under \ac{UHV} conditions for 7 days. The sample was again
cooled to 80~K and another series of spectra was taken. This process was
repeated two more times resulting in four sets of spectra, immediately after
metal deposition, 7 days after deposition, 14 days after deposition, and 21 days
after deposition. All of the spectra recorded were fit to the simplified form of
the BEEM model $I_B\propto(V_{t}-\phi_{b})^{n}$ to determine the Schottky
barrier height, where $I_B$ is the \ac{BEEM} current, $\phi_b$ is the Schottky
barrier, $V_t$ is the tip bias, and $n$ is a fitting exponent of $5/2$, given by
the \ac{PL} fitting model~\cite{prietsch:prl:66:2511}. The fits were preformed
by linearizing the spectra and using standard linear regression, which returned
the Schottky barrier and corresponding $\textit{R}^2$ value, indicating the
quality of the fit. If the fitting model cannot fit the spectrum to an
$\textit{R}^2$ value above 0.6 the spectrum is considered unable to be fit.

\section{Results}

\begin{figure}
\includegraphics{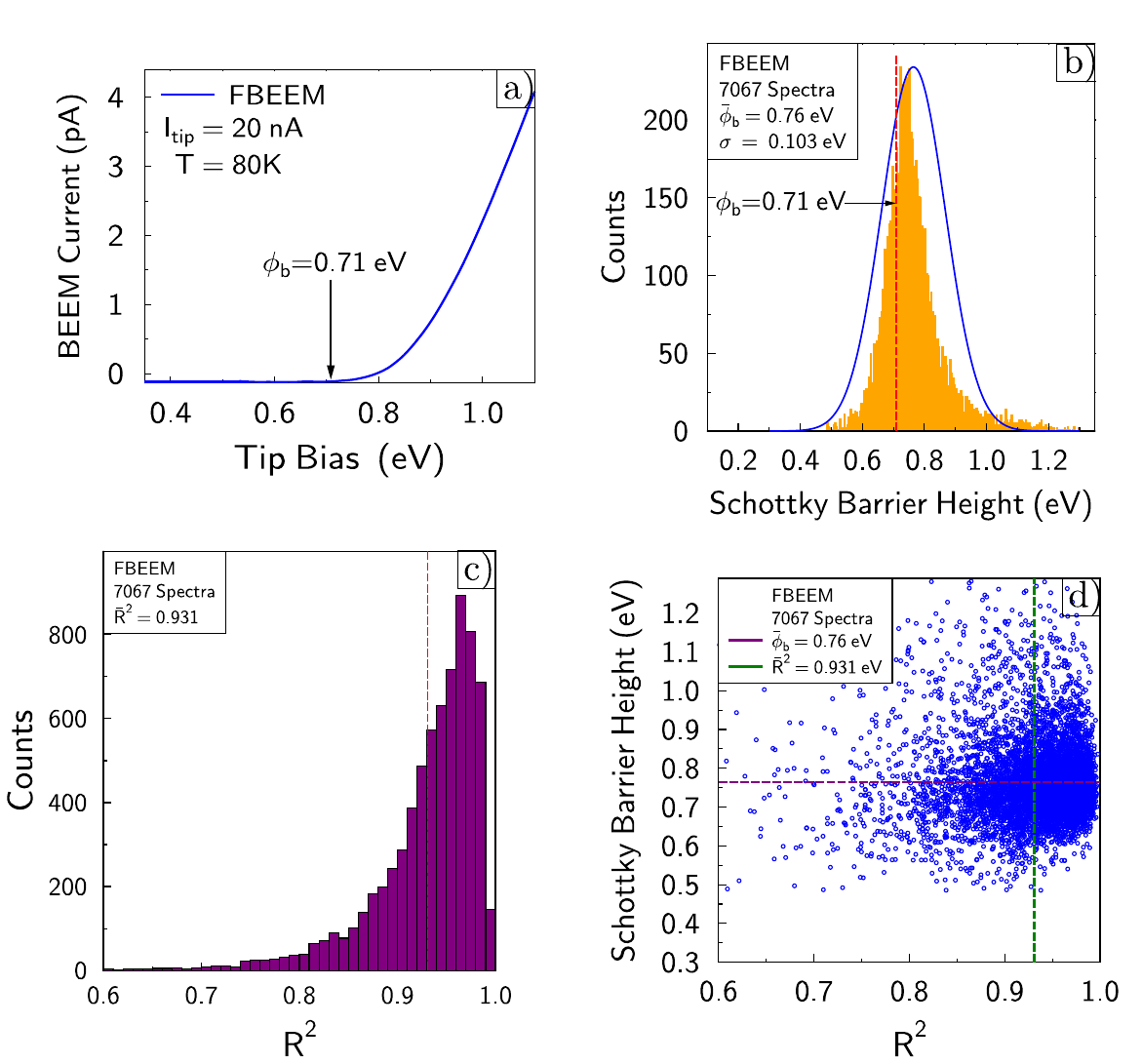}
\centering
\caption{(color online) a) \ac{BEEM} spectrum for initial set of spectra taken
immediately after metal deposition. b) Histogram of the 7,225 Schottky barriers
with a mean barrier height of 0.76~eV and dotted red line indicating the 0.71~eV
Schottky barrier from the fit to the average spectrum. The Histogram has a
standard deviation of 103 meV. c) Histogram of the $\textit{R}^2$ values with an
average $\textit{R}^2$ of 0.93. d) $\textit{R}^2$ vs Schottky barrier height
plot for the same 7,225 spectra shown in (b) with an average $\textit{R}^2$
value of 0.93.}
\label{histograms_scatter}
\end{figure}

The average spectrum from the initial \ac{BEEM} map along with the Schottky
barrier height obtained from the fit is displayed in
Fig.~\ref{histograms_scatter}(a). A histogram of all the Schottky barrier
heights from each individual spectrum along with a line indicating the fit to
the average spectrum is displayed in Fig.~\ref{histograms_scatter}(b). A
Gaussian distribution is drawn from the mean of the distribution, $\bar
\phi$~=~0.76~eV and its standard deviation, $\sigma$~=~103~meV. The Schottky
barrier height distribution has a tail to higher energies ($>$0.9~eV). A
histogram of the $\textit{R}^2$ values is displayed in
Fig.~\ref{histograms_scatter}(c) with 78$\%$ of the spectra with
$\textit{R}^2$$>$0.9 and a mean $\textit{R}^2$ value of 0.93. A scatter of the
Schottky barrier heights vs the $\textit{R}^2$ value is shown in
Fig.~\ref{histograms_scatter}(d). The plot shows that high $\textit{R}^2$ values
are obtained for the full range of Schottky barrier heights while also being
centered around the mean Schottky barrier.

\begin{figure}
\includegraphics{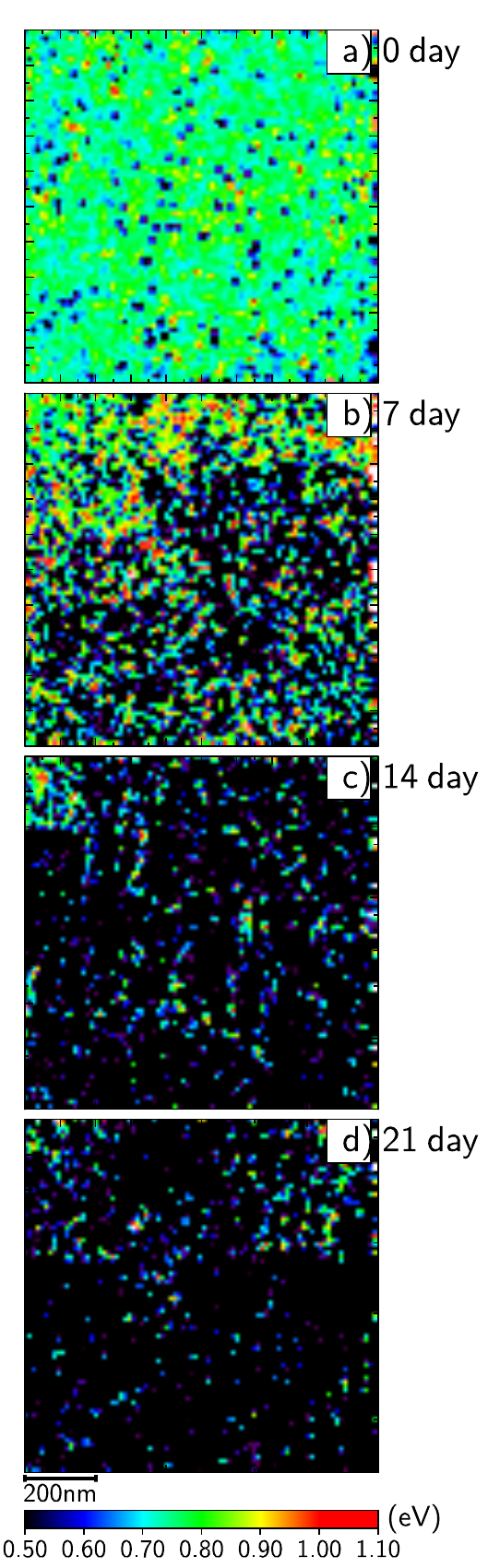}
\caption{(color online) a) Schottky barrier map of the W/n-Si(001) diode
immediately upon metal deposition. b) Schottky barrier map after seven days at
room temperature and \ac{UHV} conditions. c) Schottky barrier map after fourteen
days at room temperature and \ac{UHV} conditions. d) Schottky barrier map after
twenty-one days at room temperature and \ac{UHV} conditions.}
\label{n-maps}
\end{figure}

The Schottky barrier map acquired immediately after metal deposition, 7 days, 14
days and 21 days after deposition is shown in Fig.~\ref{n-maps}(a)-(d),
respectively. Immediately after deposition, the Schottky barrier map is uniform
with 98$\%$ of the spectra fit and centered around the average Schottky barrier
height of 0.76~eV. The few black spots are spectra with no fit. After 7 days,
the Schottky barrier map is no longer uniform, with 61$\%$ of the spectra fit,
$\bar \phi$~=~0.94~eV, and $\sigma$~=~155~meV. After 14 days, the Schottky
barrier map consists of 33$\%$ of the spectra which could be fit, $\bar
\phi$~=~0.89~eV, and $\sigma$~=~202~meV. After 21 days, the Schottky barrier map
contains only 27$\%$ of spectra which are able to be fit, $\bar \phi$~=~0.88~eV
and $\sigma$~=~221~meV.

\begin{figure}
\includegraphics{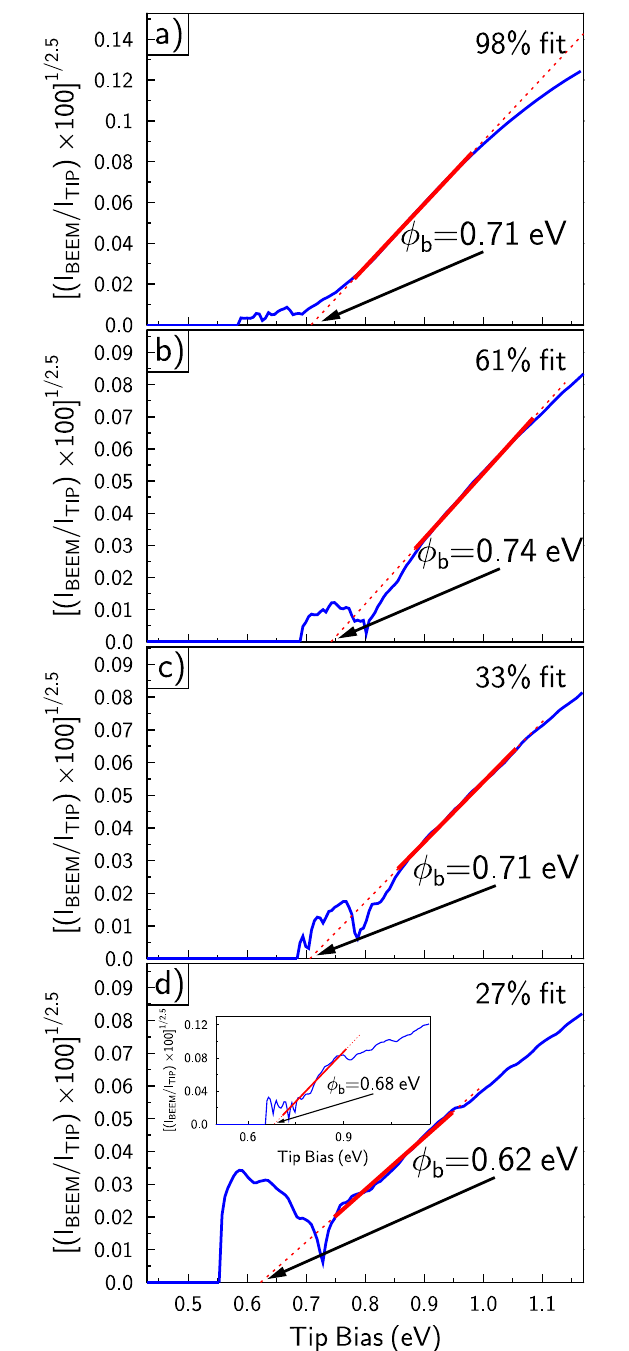}
\caption{(color online) a) Average spectrum from the initial set of \ac{BEEM}
spectra with a Schottky barrier height of 0.71~eV and 98$\%$ of the spectra able
to be fit. b) Average spectrum from the +7 day set of \ac{BEEM} spectra with a
Schottky barrier height of 0.74~eV and 61$\%$ of the spectra able to be fit. c)
Average spectrum from the +14 day set of \ac{BEEM} spectra with a Schottky
barrier height of 0.71~eV and 33$\%$ of the spectra able to be fit. d) Average
spectrum from the +21 day set of \ac{BEEM} spectra with a Schottky barrier
height of 0.62~eV and 27$\%$ of the spectra able to be fit. Inset is the average
spectrum containing only spectra with an $\textit{R}^2$ value greater than 0.95,
with a Schottky barrier height of 0.68~eV.}
\label{n-fits}
\end{figure}

The set of spectra for each Schottky barrier map were average together to form
one representative spectrum, and are shown linearized with their corresponding
fit in Fig.~\ref{n-fits}(a)-(d). The average spectra taken immediately after
deposition is fit to a Schottky barrier height of 0.71~eV and an $\textit{R}^2$
value $>$ 0.999. The 7 day sample is fit to a Schottky barrier of 0.74~eV and an
$\textit{R}^2$ value of 0.997. The 14 day sample is fit to a Schottky barrier of
0.71~eV and an $\textit{R}^2$ value of 0.997. The 21 day sample is fit to a
Schottky barrier of 0.62~eV and an $\textit{R}^2$ value of 0.98. Inset to
Fig.~\ref{n-fits}(d) is the average of all the spectra with an $\textit{R}^2$
$>$ 0.95 and is fit to a Schottky barrier of 0.68~eV.

\begin{figure}
\includegraphics{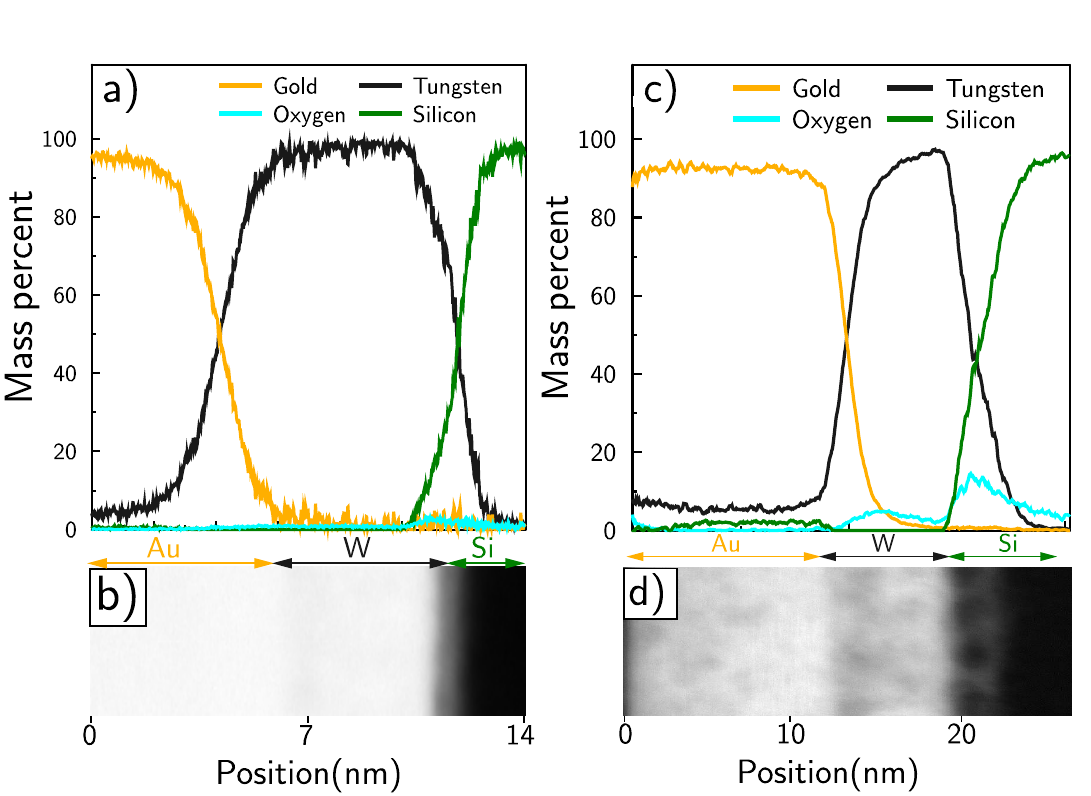}
\centering
\caption{(color online) a) \ac{EDX} line scan of the W/n-Si(001) Schottky diode
immediately after metal deposition, showing atomic percentages of Au, W, O, Si.
b) \ac{TEM} image of \ac{EDX} scan in (a). c) \ac{EDX} line scan of the W/n-
Si(001) Schottky diode 21 days after metal deposition, also showing atomic
percentages of Au, W, O, Si. d) \ac{TEM} image of the \ac{EDX} scan in (c).}
\label{metrology}
\end{figure}

The results of the \ac{TEM} and \ac{EDX} measurements for the interface
immediately after deposition are shown in Fig.~\ref{metrology}(a)~$\&$~(b). The
mass percentages of tungsten, gold, silicon, and oxygen are plotted in
Fig.~\ref{metrology}(a) with an image across the interface shown in
Fig.~\ref{metrology}(b). The \ac{EDX} and \ac{TEM} measurements of the diode in
Fig.~\ref{metrology}(a)~$\&$~(b) are taken within 24 hours of metal deposition.
The percentage of gold at the tungsten to silicon interface is negligible as
well as the percentage of oxygen. The intermixed region is $\sim$1.4~nm thick
and composed entirely of silicon and tungsten. The \ac{EDX} and \ac{TEM}
measurements of the diode 21 days after the metal deposition are shown in
Fig.~\ref{metrology}(c)~$\&$~(d). The mass percentages of tungsten, gold,
silicon, and oxygen are plotted in Fig.~\ref{metrology}(c) with an image across
the interface shown in Fig.~\ref{metrology}(d). The intermixed region has
increased it's thickness to $\sim$3.8~nm while the percentage of gold is still
negligible. There also appears to be a small oxygen concentration at the
interface.

\begin{figure}
\includegraphics{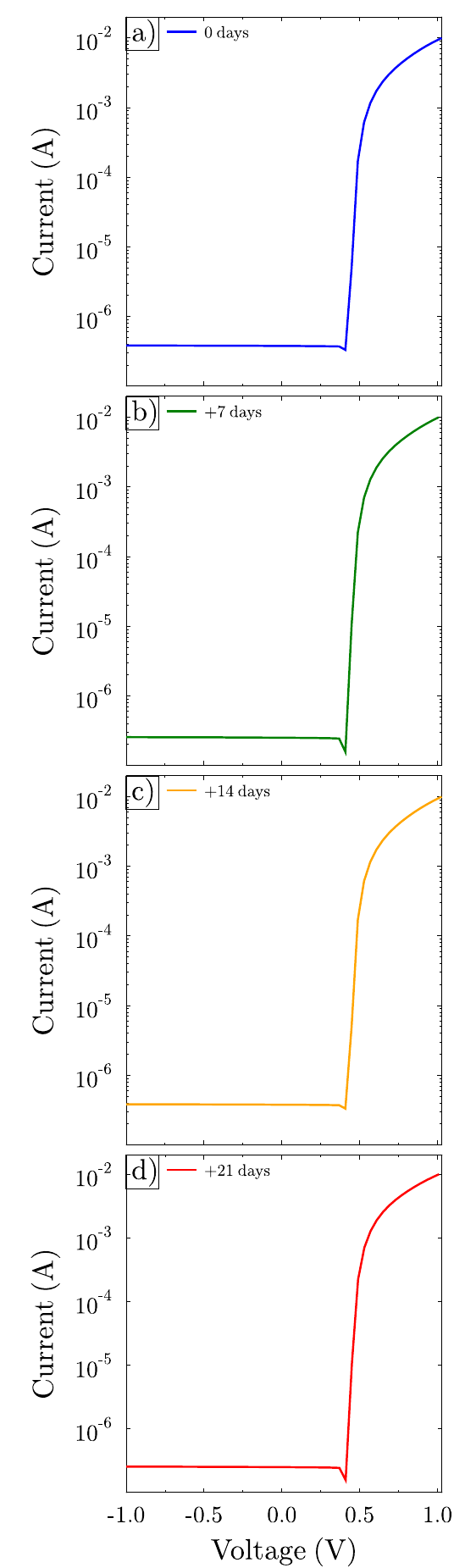}
\caption{(color online) a) Two point I-V curve for the W/n-Si(001) diode
immediately after metal deposition showing rectification. Measurement taken at
80~K without any ambient light. b) Two point I-V curve taken on the +7 day
sample showing little change in rectifying current compared to (a). c) Two point
I-V curve taken on the +14 day sample showing little change in rectifying
current. d) Two point I-V curve taken on the +21 day sample again showing little
change in rectifying current.}
\label{IV}
\end{figure}

Two point \ac{IV} curves acquired \textit{in situ} show nearly identical
rectification of the diode for all time periods throughout the study and are
displayed in Fig.~\ref{IV}(a)-(d). All of the \ac{IV} curves are taken
immediately before \ac{BEEM} measurements.

\section{Discussion}

The Schottky barrier height of the average spectra is approximately 50~meV lower
than the mean of the distribution, arising from the broad distribution Schottky
barrier heights obtained from fitting the thousands od individual spectra.
Averaging a collection of spectra with a distribution of onset biases results in
an average spectra favoring a lower onset which is picked out by the fitting
routine. The distribution of onset voltages also explains the $\sim$70~meV
difference between the Schottky barrier height and the start of the fit (solid
red line) that is observed in linearized spectra shown in
Fig.~\ref{histograms_scatter}(a). The Schottky barrier obtained from the fit to
the average spectra has been shown to be the representative barrier height of
the sample, but clearly does not capture the complicated nature of the
electrostatics that is present at the interface. The broad distribution of
barrier heights is indicative of an electrostatically varying interface arising
from the structural disorder of the tungsten silicon intermixing.

Although the majority of spectra fit to a Gaussian distribution centered around
the mean Schottky barrier height (0.76~eV), there are locations at the interface
with higher barrier heights ($\phi_{b}~>$~0.9~eV), several standard deviations
above the mean.  The $\textit{R}^2$ value of these higher barrier heights are of
high quality, $>$~0.9, as displayed in Fig.~\ref{histograms_scatter}(d). These
higher barrier heights are spatially grouped together and are distributed
randomly over the interface as seen in  Fig.~\ref{n-maps}(a).  A locally higher
barrier height can arise due to a change in the interface, such as a defect, or
the presence of a foreign material at the interface such as a oxide cluster.  In
addition, a physical defect that increases the elastic scattering can also give
rise to a locally higher barrier due to the narrow acceptance cone of parallel
momentum states just above the barrier.

A dramatic change is observed in measured electrostatics of the interface when
the \ac{BEEM} measurements are taken over the course of 21 days. The marked
decrease in the percentage of spectra that can be fit to a Schottky barrier
height indicates that structure of the interface is changing. The decrease in
the Schottky barrier height for the 21 day sample is caused by averaging
numerous spectra with little or no signal increasing the noise and making it
difficult for the fitting routine to obtain a good fit as evidenced in the lower
$R^2$ value.  To reduce the amount of noise in the spectra for the 21 day sample
an average spectra was created only from the spectra that had fits with an
$\textit{R}^2$ value above 0.95, which resulted in a higher barrier height of
0.68~eV, closer to the as-deposited sample. The amount of subthreshold current
in this spectra is also reduced. This selective averaging method can be utilized
as a powerful noise reduction method for future BEEM studies and maybe useful
for future BEEM studies where there is a low amount of transmission current.

The formation of the interfacial silicide was not accelerated by sample
annealing as the sample was left at room temperature under UHV conditions over
the 21 day period of study. Upon metal deposition the interface is comprised of
tungsten and silicon, with a 1.4~nm region of uniformly intermixed tungsten and
silicon as seen in the \ac{TEM} results displayed in Fig.~\ref{metrology}(b).
This initial intermixing is accelerated due to the unintentional heating during
the deposition process. After 21 days the interface is dramatically different,
with a much less uniform intermixed region of tungsten and silicon that is
approximately 3.8~nm thick.  The small oxygen peak at the interface is
attributed to the \ac{TEM} sample fabrication process, as the sample was not
exposed to oxygen during BEEM measurements. At no point during testing was gold
present at the interface. The 118~meV increase in the standard deviation and
drastic reduction in spectra fit over the 21 days is attributed to these
observed changes in the tungsten silicon interface. The attenuation of the
ballistic electrons increases as the intermixed region increases in both
thickness and disorder. This will result in numerous spectra with little or no
current and hence, no detectible Schottky barrier. The amount of \ac{BEEM}
transmission in the initial sample was only 0.015$\%$ at a tip bias of 1.5~V
bias, so a small change to the interface and it's thickness can induce drastic
changes to the \ac{BEEM} spectra. Interestingly, the two point I-V curves
displayed in Fig.~\ref{IV} show little to no variation in shape even with the
observed changes in the interface and BEEM measurements. This demonstrates the
utility of the BEEM technique to map changes in the electrostatics of a buried
interface to nanoscale dimensions that go unnoticed with conventional
macroscopic spectroscopy methods.

\section{Conclusion}

Utilizing \ac{BEEM}, the changes in the W/Si(001) interface over time were
measured and show changes to the transmission of ballistic electrons as the
interface evolves. The initial set of \ac{BEEM} spectra show an interface
visible to ballistic electrons with a Schottky barrier of 0.71~eV. Over the
course of 21 days the interface changes in structure causing the interface the
completely scatter ballistic electrons and inhibit the ability to accurately
measure a Schottky barrier. \ac{TEM} measurements show the interfacial silicide
which forms upon deposition grows in thickness over time at room temperature. I-
V measurements taken alongside \ac{BEEM} data show no change to the electronic
structure of the W/Si diode. These results demonstrate \ac{BEEM}'s sensitivity
to changes in the interface electrostatics with nanoscale resolution that is not
possible with conventional \ac{IV} spectroscopy.

\section{Acknowledgments}

The authors acknowledge the support of the Semiconductor Research Corporation,
Center for advanced Interconnect Science and Technology, the National Science
Foundation Grant DMR-1308102, and SEMATECH.

\pagebreak

\pagebreak

\pagebreak

\pagebreak

\pagebreak

\pagebreak

\pagebreak

\end{document}